\newcommand{\tstern}{\ensuremath{T_2^*} }

\newcommand*{\fg}[1]{Fig.\thinspace\ref{#1}}

\documentclass[aps,prb,twocolumn,superscript,superscriptaddress]{revtex4-1}

\usepackage{graphicx}
\usepackage{dcolumn}

\usepackage{bm}
\usepackage[english]{babel}

\usepackage{amsmath}
\usepackage{multirow}
\usepackage{upgreek}
\usepackage{color}

\begin{document}

%

\title{Coherent Electron Zitterbewegung}


%

\author{I. Stepanov}\affiliation{2nd Institute of Physics and JARA-FIT, RWTH Aachen University, D-52074 Aachen, Germany}
\author{M. Ersfeld}\affiliation{2nd Institute of Physics and JARA-FIT, RWTH Aachen University, D-52074 Aachen, Germany}
\author{A. V. Poshakinskiy}\affiliation{Ioffe Institute, 194021 St Petersburg, Russia}
\author{M. Lepsa}\affiliation{Peter Gr\"{u}nberg Institut (PGI-9) and JARA-FIT, Forschungszentrum J\"{u}lich GmbH, D-52425 J\"{u}lich, Germany}
\author{E. L. Ivchenko}\affiliation{Ioffe Institute, 194021 St Petersburg, Russia}
\author{S. A. Tarasenko}\affiliation{Ioffe Institute, 194021 St Petersburg, Russia}
\author{B. Beschoten}
  \thanks{ e-mail: bernd.beschoten@physik.rwth-aachen.de}
 \affiliation{2nd Institute of Physics and JARA-FIT, RWTH Aachen University, D-52074 Aachen, Germany}

\date{\today}

\maketitle
%
\textbf{Zitterbewegung is a striking consequence of relativistic quantum mechanics which predicts that free Dirac electrons exhibit a rapid trembling motion even in the absence of external forces\cite{Original_Zitterbewegung, Zawadzki_ZitterbewegungReview}. The trembling motion of an electron results from the interference between the positive and the negative-energy solutions of the Dirac equation, separated by one MeV, leading to oscillations at extremely high frequencies which are out of reach experimentally. Recently, it was shown theoretically that electrons in III-V semiconductors are governed by similar equations in the presence of spin-orbit coupling\cite{Schliemann_PhysRevLett.94.206801, Schliemann_PhysRevB.73.085323, Biswas_2012}. The small energy splittings up to meV result in Zitterbewegung at much smaller frequencies which should be experimentally accessible as an AC current. Here, we demonstrate the Zitterbewegung of electrons in a solid. We show that coherent electron Zitterbewegung can be triggered by initializing an ensemble of electrons in the same spin states in strained $n$-InGaAs and is probed as an AC current at GHz frequencies. Its amplitude is shown to increase linearly with both the spin-orbit coupling strength and the Larmor frequency of the external magnetic field. The latter dependence is the hallmark of the dynamical generation mechanism of the oscillatory motion of the Zitterbewegung. Our results demonstrate that relativistic quantum mechanics can be studied in a rather simple solid state system at moderate temperatures. Furthermore, the large amplitude of the AC current at high precession frequencies enables ultra-fast spin sensitive electric read-out in solids.} \\

%

Early on, Schr\"{o}dinger realized that a free relativistic electron, described by the Dirac Hamiltonian, is predicted to oscillate in space resulting from the interference of the positive and the negative-energy solutions of the Dirac equation\cite{Original_Zitterbewegung, Zawadzki_ZitterbewegungReview}. The frequency of this oscillation is of the order of $f = 2mc^2/h \sim 10^{20}$\,Hz while the amplitude given by the Compton wavelength $\lambda_C \sim 10^{-13}$\,m is extremely small. Both are inaccessible with present experimental techniques. Moreover, the determination of electron's position with precision better than $\lambda_C$ meets fundamental obstacles\cite{Huang1952}.
Recently, it was suggested that Zitterbewegung is not limited to free electrons but is a common feature of systems with a gapped or spin-split energy spectrum exhibiting a formal similarity to the Dirac Hamiltonian\cite{Zawadzki_ZitterbewegungReview, Winkler_PhysRevB.75.205314}. An experimental simulation of Zitterbewegung was demonstrated by a single $^{40}$Ca$^+$ ion in a linear Paul trap\cite{Gerritsma2010} and by a Bose-Einstein condensate\cite{Qu2013,ZB-Bose-Einstein} set to behave as a one-dimensional Dirac particle. Electron Zitterbewegung has not yet been demonstrated experimentally, but it has been predicted in a variety of systems with a $\bm k$-linear energy dispersion such as graphene\cite{Katsnelson2006,ZB_graphene_magnetic_field}, topological insulators\cite{ZB_TI} and in III-V semiconductor nanostructures with spin-orbit interaction\cite{Schliemann_PhysRevLett.94.206801,Biswas_2012, Winkler_PhysRevB.75.205314,Manchon.2015}. The energy splittings in solid state systems are many orders of magnitude smaller than those of free Dirac particles, resulting in relatively low frequencies which should be experimentally accessible.

Here, we probe the Zitterbewegung of electrons in $n$-type InGaAs as an AC electric current. We trigger the coherent electron Zitterbewegung (CEZ) by optical initialization of an ensemble of electron spins in the same spin states and control the frequency of electron oscillations in real space by tuning the Larmor spin precession frequency in an external magnetic field $\bm B$.

Microscopically, the Zitterbewegung under study originates from the fact that, in quantum mechanics, the electron velocity is not a conserved quantity in the presence of spin-orbit interaction (SOI), which can be readily demonstrated. Considering $\bm k$-linear spin-orbit interaction and the Zeeman splitting of electron spin states in the magnetic field $\bm B \parallel x$ results in the effective electron Hamiltonian

\begin{equation}
H = \frac{\hbar^2k^2}{2m^*} + \sum_{i,j = x,y} \beta_{ij} \sigma_{i} k_{j} + \frac{\hbar}{2}\omega_L\sigma_x \:,
\label{OurHamiltonian}
\end{equation}

where $m^*$ is the effective mass, $\beta_{ij}$ are the SOI constants originating from Rashba and Dresselhaus SOI in (001)-oriented structures, $\sigma_i$ are the Pauli spin matrices, $\omega_L = g \mu_B B_x /\hbar$ is the Larmor frequency, $g$ is the effective $g$-factor, $\mu_B$ is the Bohr magneton, $x$ and $y$ are the in-plane axes, with $z \parallel [001]$.
Besides the first term of spin-independent kinetic energy, the effective Hamiltonian~\eqref{OurHamiltonian}
mimics the Dirac Hamiltonian with all essential elements: it comprises the SOI term which provides a $\bm k$-linear coupling of the states and the Zeeman term which opens a gap at $\bm k=0$ and plays a role of the mass term in the Dirac Hamiltonian. The Zeeman splitting is many orders of magnitude smaller than the electron-positron gap suggesting the emergence of Zitterbewegung at much lower, here GHz, frequencies.

\begin{figure*}[tbp]
\includegraphics{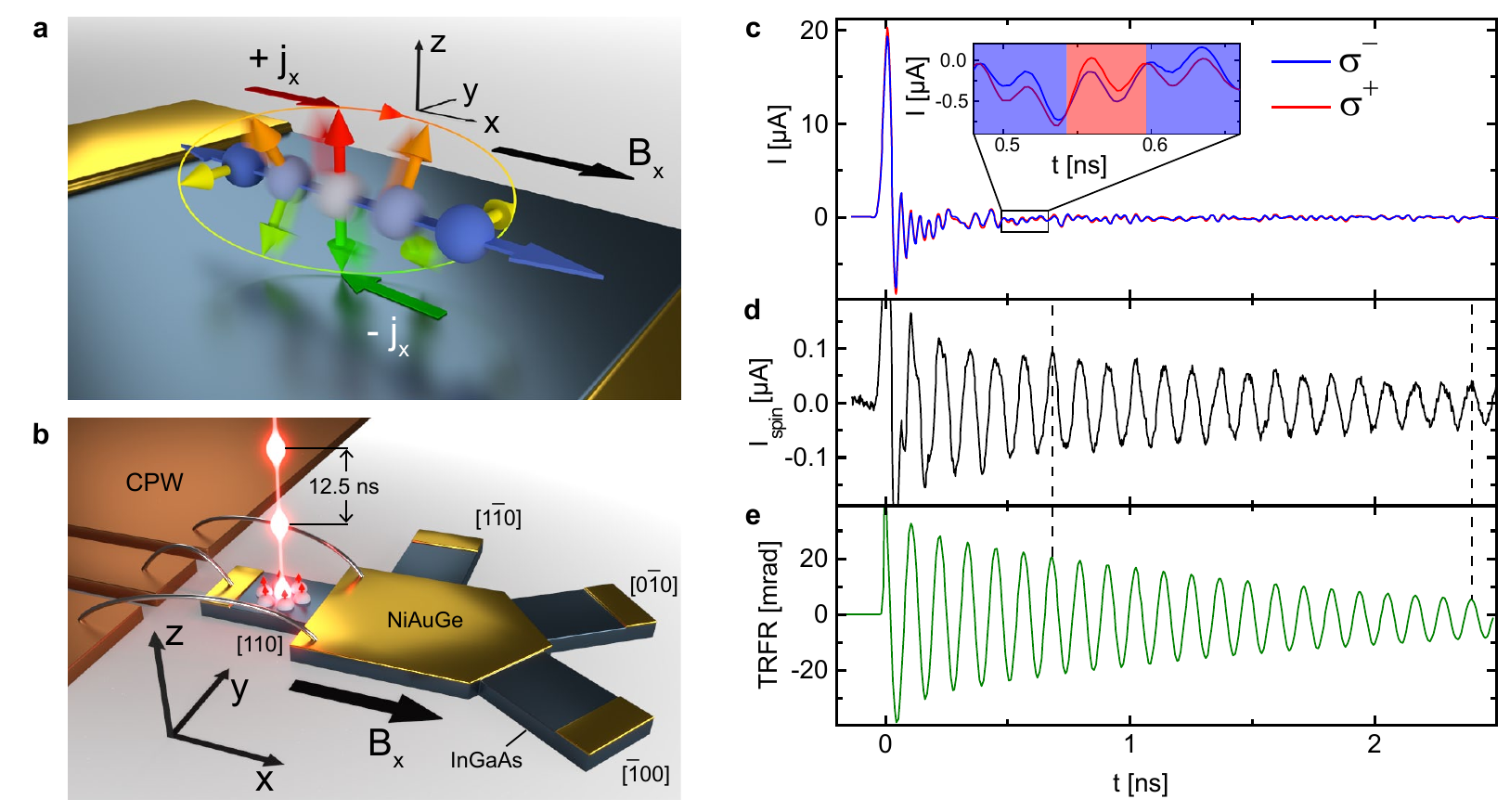}
\caption{\label{fig1} Time-resolved detection of coherent electron Zitterbewegung as a high frequency AC current. \textbf{a}, Illustration of coherent electron Zitterbewegung emerging in an external magnetic field $B_x$. Electron spin precession yields a periodic displacement of the electrons along the $x$-direction which results in an AC current oscillating at the Larmor frequency. \textbf{b}, Sample geometry and pulsed optical spin excitation. The electron spin ensemble is excited along the $z$ direction by circularly polarized ps laser pulses with a pulse repetition time of 12.5 ns. The high frequency current of the coherent electron Zitterbewegung in InGaAs is transferred into the coplanar waveguide (CPW) by three bond wires and detected by a phase-triggered sampling oscilloscope. Four different crystal directions can be contacted. \textbf{c}, Time-resolved traces of the current along the $[110]$ crystal direction recorded after pulsed optical excitation for $\sigma^+$ and $\sigma^-$ polarizations at $B_x = 1$\,T and $T = 50$~K. The inset shows a close-up of both current traces. \textbf{d}, Time-dependent AC current $I_{\rm spin} = [I(\sigma^+)-I(\sigma^-)]/2$ as determined from both traces in panel b demonstrating the spin precession driven coherent electron Zitterbewegung. \textbf{e}, Time-resolved Faraday rotation measurement of electron spin precession under identical experimental conditions. }
\end{figure*}

Calculating the electron velocity operator $v_x = (i/\hbar) [H,x]$, where $[A,B]=AB-BA$ is the commutator of the operators $A$ and $B$,
and then the acceleration operator $\dot{v}_x = (i/\hbar) [H,v_x]$, we obtain to first order in SOI

\begin{equation}
\dot{v}_x = -\frac{\beta_{yx}}{\hbar}\omega_L\sigma_z \:.
\label{velocity_operator}
\end{equation}

Equation~\eqref{velocity_operator} describes the coupling of the electron acceleration $\dot{v}_x$ and its spin projection $s_z =\sigma_z/2$ which, in turn, satisfies the operator equations $\dot{s}_z = \Omega_L s_y$ and $\dot{s}_y = -\Omega_L s_z$. Solution of the above dynamic equations yields

\begin{equation}
v_x(t) =  v_x(0) + \frac{2\beta_{yx}}{\hbar}\left[s_y(0) (\cos\omega_Lt-1) -  s_z(0) \sin\omega_Lt \right]
 \:.\\
\label{x_heis}
\end{equation}

The oscillating contribution to the velocity represents the Zitterbewegung of ballistic electrons originating from the interference of spin states separated by the Zeeman gap.
From Eq.~\eqref{x_heis} we estimate the amplitude of the trembling as 20\,nm.

In the samples we study, the electron transport is diffusive rather than ballistic and we deal with an ensemble of about $10^8$ electrons.
Averaging Eq.~\eqref{velocity_operator} over the electron ensemble and including electron scattering by phonons or static defects, which acts as a ``friction force'' slowing down the electron velocity with  no considerable effect on the spin dynamics, we derive the relation to the average electron velocity $\bar{v}_x$ and the average spin projection $\bar{s}_z$
\begin{eqnarray}
\dot{\bar{v}}_x(t) = -2\frac{\beta_{yx}}{\hbar}\omega_L \bar{s}_z(t) - \frac{\bar{v}_x(t)}{\tau_p} \:,
\label{velocity}
\end{eqnarray}
where $\tau_p$ is the momentum scattering time. Equation~\eqref{velocity} allows for studying the electron Zitterbewegung by exploring the AC electric current density $j^{(ac)}_x(t) = e n_e \bar{v}_x(t)$ caused by the coherent trembling motion of electrons, where $e$ is the electron charge and $n_e$ is the electron density. For the case $\omega_L\tau_p\ll 1$, which is common and realized in our
experiments, solution of Eq.~\eqref{velocity} yields
\begin{eqnarray}
j^{(ac)}_x(t) = -2en_e\frac{\beta_{yx}}{\hbar} \omega_L \tau_p \bar{s}_z(t) \:.
\label{current_density}
\end{eqnarray}
In an external magnetic field the spin projection $\bar{s}_z(t)$ oscillates at the Larmor frequency $\omega_L$ and so does the electron velocity even though no driving electric field is applied. The CEZ current exhibits one distinct frequency which linearly increases with increasing magnetic field strength, despite the randomized electron momenta present in the diffusive transport regime.

Equation~\eqref{current_density} shows that CEZ emerges only during the spin precession for $\omega_L\neq0$. The temporal evolution of the spin vector $\bar{s}(t)$ during spin precession and the predicted AC current is illustrated in \fg{fig1}a. Coherent spin precession and coherent Zitterbewegung movement are in phase, i.e., the largest current along the $+x$ direction is obtained when the spin is pointing along $+z$ (red arrow). The current becomes zero at the spin rotation angle $\pi/2$ ($S_z=0$, see yellow arrow in \fg{fig1}a), changes the sign for larger angles, and again becomes strongest along the $-x$ direction for the angle $\pi$ (green arrow). To verify this notion experimentally, we use circularly polarized ps laser pulses to optically create phase-coherent spin ensembles in $n$-doped InGaAs (see \fg{fig1}b). Spin precession is recorded according to Eq.~\eqref{current_density} as a time-resolved AC current. Time-resolution provides direct access to both the amplitude and the phase of the AC current which allow us to easily distinguish spin precession-induced CEZ from other spin-driven charge currents as, for example, expected from the spin-galvanic or inverse spin Hall effects\cite{Nature417_Ganichev2002_Spin-GalvanicEffect, Schmidt_APL2015}. We use $n$-InGaAs as it exhibits both Dresselhaus SOI due to strain by lattice mismatch to the subjacent SI-GaAs substrate and Rashba-like SOI due to partial strain relaxation\cite{PhysRevB.72.115204}. By analyzing the AC current strength along different crystal directions, we can directly verify that $j^{(ac)}_x(t)$ depends on the SOI strength $\beta_{yx}$ as predicted in Eq.~\eqref{current_density}.

Circularly polarized ps laser pulses with photon energies of 1.41~eV near the fundamental band edge of InGaAs are used to trigger an initial spin polarization $S_z(0)$ at $t=0$~ns pointing along $+z$ and $-z$ directions for $\sigma^+$ and $\sigma^-$ polarizations, respectively\cite{Meier_Optical_Orientation}. Spins precess about an external magnetic field $\bm B \parallel x$ and the generated current is probed through high frequency contacts (see \fg{fig1}b) \cite{PhysRevLett.109.146603,apl1.4864468}. The time-dependent current signal measured at $T=50$~K along the $\bm B$ field direction ($B_x=1$\,T) that is oriented parallel to the [110] crystal axis is shown in \fg{fig1}c for $\sigma^+$ and $\sigma^-$ excitation. For short $t$, it is dominated by a spin-independent background which decays on the electron-hole recombination time of about 100 ps and is followed by some circuit ringing due to internal reflections in the transmission line. Spin precession is already visible in this data as subtle periodic differences between both curves which is shown in the close-up region in \fg{fig1}c. We can enhance the visibility of the spin-dependent signal by plotting $I_{\rm spin}=[I(\sigma^+)-I(\sigma^-)]/2$ in \fg{fig1}d. An oscillating current is clearly visible. It extends over several ns after the laser-induced non-equilibrium electron-hole population has recombined indicating that the spin angular momentum has been transferred to the resident electron ensemble. To unambiguously demonstrate that the current oscillations result from Larmor precession, we show time-resolved pump-probe Faraday rotation (TRFR) data in \fg{fig1}e which have been obtained under identical experimental conditions. The time-resolved oscillating current exhibits the same Larmor precession frequency and temporal decay as the TRFR data showing that the same spin states are probed by both techniques. The AC current can be fitted by an exponentially decaying cosine function

\begin{equation}
I_{\rm spin}(t) = I_0\cdot\exp\left(-\frac{t}{\tstern}\right)\cdot\cos\left(\omega_L t+\phi\right),
\label{Oszillation}
\end{equation}

to extract the current amplitude $I_0$, the spin dephasing time $\tstern$ and the respective phase $\phi$, with the Larmor frequency
$\omega_L$ determined by the electron g-factor $g=0.62$. As predicted for the spin-driven CEZ in Eq.~\eqref{current_density}, the AC current in \fg{fig1}d is in phase ($\phi = 0^{\circ}$) with the $z$-component of the spin polarization which is directly probed by TRFR (for comparison of phase see dotted lines in \fg{fig1}\,d and e). Moreover, the current amplitude estimated from Eq.~\eqref{current_density} for the sample parameters determined independently (see methods and supplements) $I_0 \sim 0.2$~$\mu$A is in agreement with the experimental data.

%
\begin{figure}[tbp]
\includegraphics{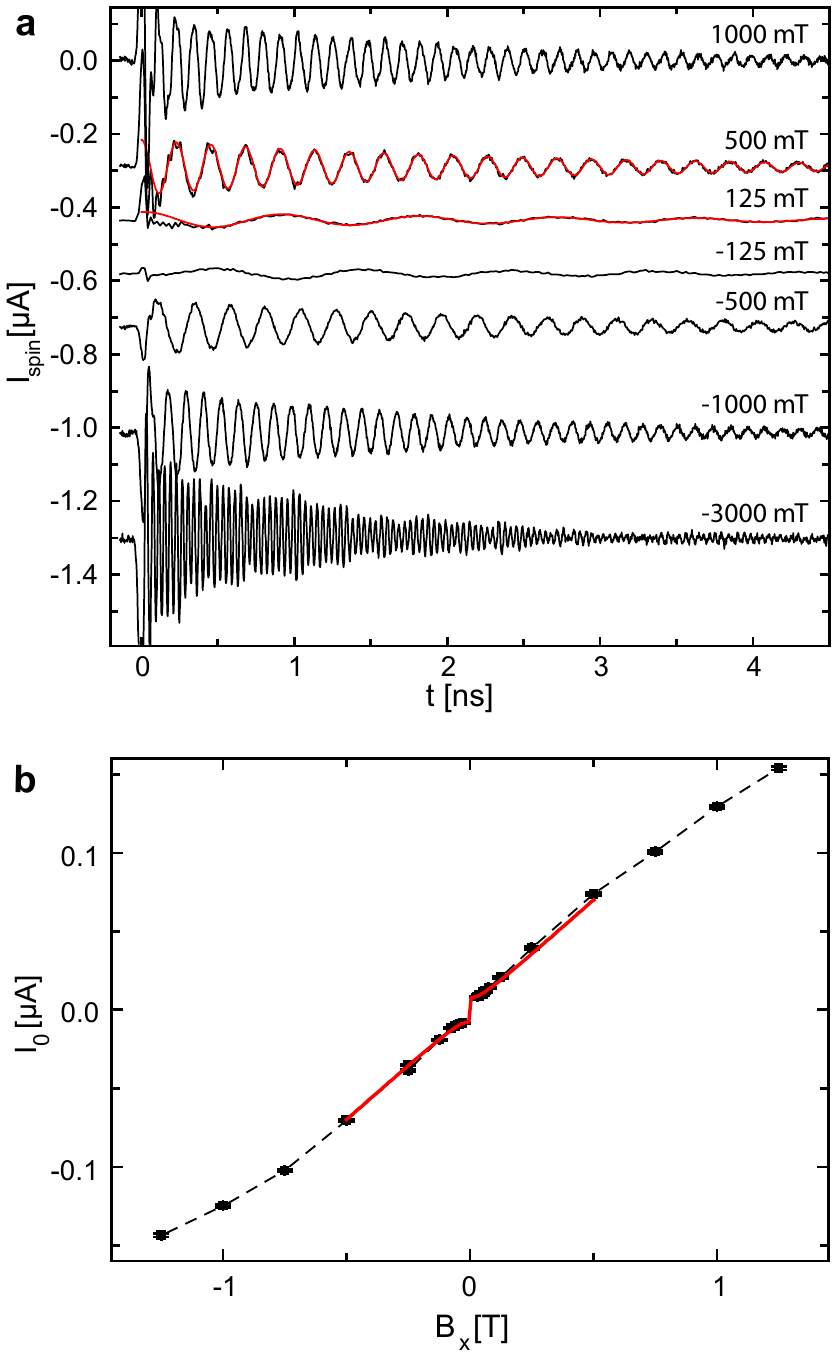}
\caption{ \label{fig2} Magnetic field dependence of spin precession driven coherent Zitterbewegung. \textbf{a}, Time-resolved AC current along the $[1 1 0]$ direction in InGaAs at various magnetic fields. Larmor precession frequency and magnitude of AC current increase with increasing magnetic field strength while the sign of the AC current reverses when reversing the magnetic field direction. \textbf{b}, Amplitude $I_0$ of AC current vs. applied magnetic field $B_x$. The linear increase of $I_0$ with $B_x$ indicates a strong CEZ, with a weak SGE visible at small $B_x$. The red solid line is a fit to Eq.~\eqref{Separation1}.}
\end{figure}

To further explore the dependence of the spin-driven AC current on $B_{x}$, we show a series of CEZ measurements at selected magnetic fields in \fg{fig2}a.
It is apparent that both the frequency $\omega_L$ and the amplitude $I_0$ increase with increasing  magnetic field strength (see also \fg{fig2}b). At large magnetic fields of $B_{x} = \pm3$\,T the spin precession frequency reaches $f=26$\,GHz (see \fg{fig2}a). In contrast to TRFR, where the signal amplitude is a measure of the total spin density and is independent of the precession frequency, the amplitude $I_0$ of CEZ is determined by the Larmor frequency which increases linearly with the magnetic field strength $B_{x}$, see Eq.~\eqref{current_density}. When reversing the magnetic field direction, the spin precession direction changes from clockwise to counterclockwise explaining the sign reversal of $I_0$ for negative magnetic fields.\\

We note that there is a deviation from the expected linear dependence of $I_0$ on $B_x$ from the CEZ theory at small magnetic fields where the Larmor precession slows down and spin dynamics becomes dominated by spin relaxation which is not included in the above equations. The processes of spin relaxation of polarized electrons can also give rise to the generation of an electric current, a phenomenon known as the spin-galvanic effect (SGE)\cite{Nature417_Ganichev2002_Spin-GalvanicEffect}. Its current emerges in the third order in the SOI constant and is driven by the in-plane spin component in (001)-oriented structures while its amplitude $I_{\rm SGE}$ does not depend on $B_{x}$\cite{PhysRevB.68.081302, PSSB:PSSB201350261, Ivchenko_JETP_Letters_1989, Ivchenko_JETP_1990}. Thus, CEZ and SGE can easily be distinguished in time-resolved experiments (see supplements). Because of the $\pi/2$ phase difference between the SGE and CEZ currents their amplitudes, $I_{\rm SGE}$ and $I_{\rm CEZ} = K_{\rm CEZ} B_{x}$, respectively, add quadratically to the AC current amplitude

\begin{equation}
I_0(B_x)=\sqrt{I_{\rm SGE}^2+(K_{\rm CEZ} B_{x})^2}, \label{Separation1}\\
\end{equation}

where $K_{\rm CEZ}$ is the strength of CEZ. This equation describes well the experimental data including the small magnetic field range. At higher precession frequencies we observe a slight deviation from the expected linear behavior due to the limited high frequency band-width of our device (3db point at 13 GHz).

\begin{figure}[tbp]
\includegraphics{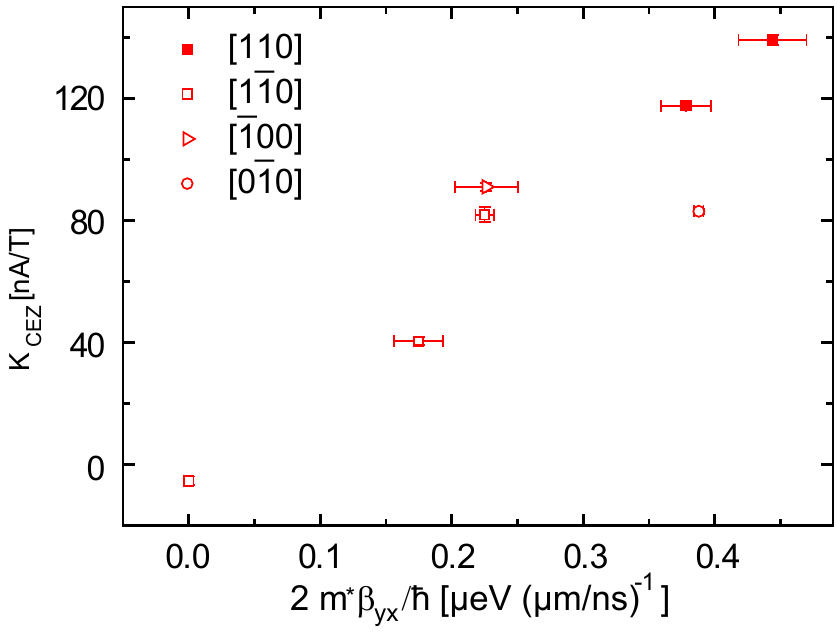}
\caption{\label{fig3} Dependence of coherent electron Zitterbewegung strength $K_{\rm CEZ}$ on the spin-orbit coupling constant $\beta_{yx}$. The data points are obtained from a number of InGaAs devices where the AC current is probed along different crystal directions. The overall linear behavior is consistent with the predicted dependence in Eq.~\eqref{current_density}. }
\end{figure}

As it follows from Eq.~\eqref{current_density}, the amplitude of the CEZ current along the magnetic field direction $\bm B \parallel x$ which is determined by $K_{\rm CEZ}$, scales linearly with the SO coupling constant $\beta_{yx}$. Due to strain and partial strain relaxation the InGaAs epilayers under study exhibit $\bm k$-linear SO coupling of both Rashba and Dresselhaus types\cite{PhysRevB.72.115204}. The resulting crystal axis anisotropy of SOI gives us a set of devices from the same wafer with different $\beta_{yx}$ (see supplements). For measuring $\beta_{yx}$ we use the TRFR pump-probe technique and apply an additional DC current $I$ which converts into an internal magnetic field $B_{\rm int}\propto I$ (refs \onlinecite{Meier_Rashba_2007, Kato2004_manipulation}). From the change in the spin precession frequency we determine $\beta_{yx}$ for each device and find both Rashba and Dresselhaus contributions. We conducted CEZ measurements on seven devices along four crystal directions (see \fg{fig1}a) and extracted the respective $K_{\rm CEZ}$ and $\beta_{yx}$ values which are summarized in \fg{fig3}. Despite a certain device-to-device variation along the same crystal direction, we clearly observe the predicted linear dependence of $K_{\rm CEZ}$ on $\beta_{yx}$ demonstrating that the SO coupling plays the key role for the emergence of CEZ.\\

%

We have shown that electrons  in III-V semiconductors with spin-orbit coupling experience an inherent trembling motion of quantum-mechanical nature. The Zitterbewegung of individual electrons can be phase-synchronized by initializing the electrons in the same spin states and detected as an AC current oscillating at the Larmor frequency of several GHz. The use of pulsed optical excitation combined with time-resolved electrical detection in systems with rather small Zeeman splitting provides a promising pathway for exploring coherent Zitterbewegung in other solid state systems. Additionally, the spin-driven AC currents can be utilized as an ultrafast spin sensitive electrical probe of electron spin precession where the detection scheme is solely based on SO interaction free of any additional spin-sensitive magnetic materials. We expect that spin-orbit-driven high frequency currents can also be explored in other materials including topological insulators with inherent Dirac-like spectrum and other 2D materials with strong spin-orbit interaction and short spin lifetimes.

%
\subsection*{Methods}
The samples consist of a 500 nm thick In$_{0.07}$Ga$_{0.93}$As epilayer grown on a semi-insulating (001) GaAs wafer by molecular beam epitaxy. The epilayer is doped with Si to an electron density of \mbox{$n\sim 3\times 10^{16}$~cm$^{-3}$} in order to allow for long spin dephasing times\cite{PRL80_Kikkawa1998_ResonantSpinAmplificationinN-TypeGaAs,2010_Schmalbuch}. The large thickness of the In$_{0.07}$Ga$_{0.93}$As epilayer introduces a gradient in the strain caused by the lattice mismatch to the underlying GaAs substrate. Ge/Ni/Au contacts, defined by chemical wet etching, were alloyed into the InGaAs layer to provide ohmic contacts. The devices have an InGaAs area of $180 \times 200$\,$\upmu $m$^2$. They are cooled down to 50~K in a helium bath cryostat, providing long enough spin dephasing time $\tstern$ while the nuclear polarization effects are negligible. Electrons are predominantly scattered at static defects with a momentum relaxation time of $\tau_p=143$~fs as determined by Hall effect measurements. The momentum relaxation time $\tau_p$ is many orders of magnitude shorter than the spin dephasing time \tstern that is typically on the order of 2\,ns.

The devices were mounted on a circuit board with additional heat sinks and are connected to a 50\,$\Omega$ coplanar waveguide by three bond wires. The high frequency AC current signal is amplified by 35\,dB and recorded by a 50~GHz sampling oscilloscope. The current is calculated by dividing the measured voltage by the impedance of 50~$\Omega$ and by the amplification factor. All AC current measurements were conducted along the direction of external magnetic field $\bm B$. For time-resolved measurements the oscilloscope is triggered by a fast photo-diode illuminated by the pulse train of the ps laser providing a fixed time-reference.

For optical spin orientation, we use a mode-locked Ti:sapphire laser with a pulse-width of $\approx 3$ \,ps and a pulse repetition rate of 80\,MHz. The time between subsequent pulses is much larger than the spin dephasing time and the CEZ currents induced by the pulses do not interfere. The initial spin polarization along $z \parallel [0 0 1]$ is created by pump pulses with an average power of 30\,mW with $\sigma^+$/$\sigma^-$ helicities which are set by a variable liquid crystal retarder.

For optical detection of spin precession in TRFR we use linearly polarized probe pulses with an average power of 500\,$\upmu$W. A mechanical delay is used to monitor the evolution of $S_z(\Delta t)$ by measuring the Faraday rotation angle $\Theta(\Delta t)\propto S_z(\Delta t)$ by a balanced photo-diode bridge. Both laser pulses are tuned to the Faraday resonance (879~nm at $T=50$~K) and are focused onto the center of the InGaAs area with a full width at half maximum value of 100\,$\upmu$m.

%
\subsection*{}
\bibliographystyle{naturemag}
\selectlanguage{english}
\bibliography{literature}

\begin{thebibliography}{10}
\expandafter\ifx\csname url\endcsname\relax
  \def\url#1{\texttt{#1}}\fi
\expandafter\ifx\csname urlprefix\endcsname\relax\def\urlprefix{URL }\fi
\providecommand{\bibinfo}[2]{#2}
\providecommand{\eprint}[2][]{\url{#2}}

\bibitem{Original_Zitterbewegung}
\bibinfo{author}{Schr\"odinger, E.}
\newblock \bibinfo{title}{{{\"U}ber die kr{\"a}ftefreie Bewegung in der
  relativistischen Quantenmechanik}}.
\newblock \emph{\bibinfo{journal}{Sitz. Press. Akad. Wiss. Phys.-Math.}}
  \textbf{\bibinfo{volume}{24}}, \bibinfo{pages}{418--428}
  (\bibinfo{year}{1930}).

\bibitem{Zawadzki_ZitterbewegungReview}
\bibinfo{author}{Zawadzki, W.} \& \bibinfo{author}{Rusin, T.~M.}
\newblock \bibinfo{title}{Zitterbewegung (trembling motion) of electrons in
  semiconductors: a review}.
\newblock \emph{\bibinfo{journal}{J. Phys.: Condens. Matter}}
  \textbf{\bibinfo{volume}{23}}, \bibinfo{pages}{143201}
  (\bibinfo{year}{2011}).

\bibitem{Schliemann_PhysRevLett.94.206801}
\bibinfo{author}{Schliemann, J.}, \bibinfo{author}{Loss, D.} \&
  \bibinfo{author}{Westervelt, R.~M.}
\newblock \bibinfo{title}{Zitterbewegung of electronic wave packets in {III-V}
  zinc-blende semiconductor quantum wells}.
\newblock \emph{\bibinfo{journal}{Phys. Rev. Lett.}}
  \textbf{\bibinfo{volume}{94}}, \bibinfo{pages}{206801}
  (\bibinfo{year}{2005}).

\bibitem{Schliemann_PhysRevB.73.085323}
\bibinfo{author}{Schliemann, J.}, \bibinfo{author}{Loss, D.} \&
  \bibinfo{author}{Westervelt, R.~M.}
\newblock \bibinfo{title}{Zitterbewegung of electrons and holes in {III-V}
  semiconductor quantum wells}.
\newblock \emph{\bibinfo{journal}{Phys. Rev. B}} \textbf{\bibinfo{volume}{73}},
  \bibinfo{pages}{085323} (\bibinfo{year}{2006}).

\bibitem{Biswas_2012}
\bibinfo{author}{Biswas, T.} \& \bibinfo{author}{Ghosh, T.~K.}
\newblock \bibinfo{title}{Zitterbewegung of electrons in quantum wells and dots
  in the presence of an in-plane magnetic field}.
\newblock \emph{\bibinfo{journal}{J. Phys.: Condens. Matter}}
  \textbf{\bibinfo{volume}{24}}, \bibinfo{pages}{185304}
  (\bibinfo{year}{2012}).

\bibitem{Huang1952}
\bibinfo{author}{Huang, K.}
\newblock \bibinfo{title}{On the zitterbewegung of the dirac electron}.
\newblock \emph{\bibinfo{journal}{American Journal of Physics}}
  \textbf{\bibinfo{volume}{20}}, \bibinfo{pages}{479--484}
  (\bibinfo{year}{1952}).

\bibitem{Winkler_PhysRevB.75.205314}
\bibinfo{author}{Winkler, R.}, \bibinfo{author}{Z\"ulicke, U.} \&
  \bibinfo{author}{Bolte, J.}
\newblock \bibinfo{title}{Oscillatory multiband dynamics of free particles:
  {T}he ubiquity of \textit{zitterbewegung} effects}.
\newblock \emph{\bibinfo{journal}{Phys. Rev. B}} \textbf{\bibinfo{volume}{75}},
  \bibinfo{pages}{205314} (\bibinfo{year}{2007}).

\bibitem{Gerritsma2010}
\bibinfo{author}{Gerritsma, R.} \emph{et~al.}
\newblock \bibinfo{title}{Quantum simulation of the {D}irac equation}.
\newblock \emph{\bibinfo{journal}{Nature}} \textbf{\bibinfo{volume}{463}},
  \bibinfo{pages}{68--71} (\bibinfo{year}{2010}).

\bibitem{Qu2013}
\bibinfo{author}{Qu, C.}, \bibinfo{author}{Hamner, C.}, \bibinfo{author}{Gong,
  M.}, \bibinfo{author}{Zhang, C.} \& \bibinfo{author}{Engels, P.}
\newblock \bibinfo{title}{Observation of \textit{Zitterbewegung} in a
  spin-orbit-coupled bose-einstein condensate}.
\newblock \emph{\bibinfo{journal}{Phys. Rev. A}} \textbf{\bibinfo{volume}{88}},
  \bibinfo{pages}{021604} (\bibinfo{year}{2013}).

\bibitem{ZB-Bose-Einstein}
\bibinfo{author}{LeBlanc, L.~J.} \emph{et~al.}
\newblock \bibinfo{title}{Direct observation of zitterbewegung in a
  {B}ose--{E}instein condensate}.
\newblock \emph{\bibinfo{journal}{New J. Phys.}} \textbf{\bibinfo{volume}{15}},
  \bibinfo{pages}{073011} (\bibinfo{year}{2013}).

\bibitem{Katsnelson2006}
\bibinfo{author}{Katsnelson, I.~M.}
\newblock \bibinfo{title}{Zitterbewegung, chirality, and minimal conductivity
  in graphene}.
\newblock \emph{\bibinfo{journal}{The European Physical Journal B - Condensed
  Matter and Complex Systems}} \textbf{\bibinfo{volume}{51}},
  \bibinfo{pages}{157--160} (\bibinfo{year}{2006}).

\bibitem{ZB_graphene_magnetic_field}
\bibinfo{author}{Rusin, T.~M.} \& \bibinfo{author}{Zawadzki, W.}
\newblock \bibinfo{title}{Zitterbewegung of electrons in graphene in a magnetic
  field}.
\newblock \emph{\bibinfo{journal}{Phys. Rev. B}} \textbf{\bibinfo{volume}{78}},
  \bibinfo{pages}{125419} (\bibinfo{year}{2008}).

\bibitem{ZB_TI}
\bibinfo{author}{Shi, L.-k.}, \bibinfo{author}{Zhang, S.-c.} \&
  \bibinfo{author}{Chang, K.}
\newblock \bibinfo{title}{Anomalous electron trajectory in topological
  insulators}.
\newblock \emph{\bibinfo{journal}{Phys. Rev. B}} \textbf{\bibinfo{volume}{87}},
  \bibinfo{pages}{161115} (\bibinfo{year}{2013}).

\bibitem{Manchon.2015}
\bibinfo{author}{Manchon, A.}, \bibinfo{author}{Koo, H.~C.},
  \bibinfo{author}{Nitta, J.}, \bibinfo{author}{Frolov, S.~M.} \&
  \bibinfo{author}{Duine, R.~A.}
\newblock \bibinfo{title}{New perspectives for rashba spin-orbit coupling}.
\newblock \emph{\bibinfo{journal}{Nat. Mater.}} \textbf{\bibinfo{volume}{14}},
  \bibinfo{pages}{871--882} (\bibinfo{year}{2015}).

\bibitem{Nature417_Ganichev2002_Spin-GalvanicEffect}
\bibinfo{author}{Ganichev, S.~D.} \emph{et~al.}
\newblock \bibinfo{title}{{S}pin-galvanic effect}.
\newblock \emph{\bibinfo{journal}{Nature}} \textbf{\bibinfo{volume}{417}},
  \bibinfo{pages}{153--156} (\bibinfo{year}{2002}).

\bibitem{Schmidt_APL2015}
\bibinfo{author}{Schmidt, C.~B.}, \bibinfo{author}{Priyadarshi, S.},
  \bibinfo{author}{Tarasenko, S.~A.} \& \bibinfo{author}{Bieler, M.}
\newblock \bibinfo{title}{Ultrafast magneto-photocurrents in {GaAs}:
  {S}eparation of surface and bulk contributions}.
\newblock \emph{\bibinfo{journal}{Appl. Phys. Lett.}}
  \textbf{\bibinfo{volume}{106}} (\bibinfo{year}{2015}).

\bibitem{PhysRevB.72.115204}
\bibinfo{author}{Bernevig, B.~A.} \& \bibinfo{author}{Zhang, S.-C.}
\newblock \bibinfo{title}{Spin splitting and spin current in strained bulk
  semiconductors}.
\newblock \emph{\bibinfo{journal}{Phys. Rev. B}} \textbf{\bibinfo{volume}{72}},
  \bibinfo{pages}{115204} (\bibinfo{year}{2005}).

\bibitem{Meier_Optical_Orientation}
\bibinfo{editor}{Meier, F.} \& \bibinfo{editor}{Zakharchenya, B.~P.} (eds.)
  \emph{\bibinfo{title}{Optical Orientation}} (\bibinfo{publisher}{Elsevier,
  Amsterdam}, \bibinfo{year}{1984}).

\bibitem{PhysRevLett.109.146603}
\bibinfo{author}{Kuhlen, S.} \emph{et~al.}
\newblock \bibinfo{title}{Electric field-driven coherent spin reorientation of
  optically generated electron spin packets in {InGaAs}}.
\newblock \emph{\bibinfo{journal}{Phys. Rev. Lett.}}
  \textbf{\bibinfo{volume}{109}}, \bibinfo{pages}{146603}
  (\bibinfo{year}{2012}).

\bibitem{apl1.4864468}
\bibinfo{author}{Stepanov, I.}, \bibinfo{author}{Kuhlen, S.},
  \bibinfo{author}{Ersfeld, M.}, \bibinfo{author}{Lepsa, M.} \&
  \bibinfo{author}{Beschoten, B.}
\newblock \bibinfo{title}{All-electrical time-resolved spin generation and spin
  manipulation in $n$-{InGaAs}}.
\newblock \emph{\bibinfo{journal}{Appl. Phys. Lett.}}
  \textbf{\bibinfo{volume}{104}}, \bibinfo{pages}{062406}
  (\bibinfo{year}{2014}).

\bibitem{PhysRevB.68.081302}
\bibinfo{author}{Ganichev, S.~D.} \emph{et~al.}
\newblock \bibinfo{title}{Spin-galvanic effect due to optical spin orientation
  in \textit{n}-type {GaAs} quantum well structures}.
\newblock \emph{\bibinfo{journal}{Phys. Rev. B}} \textbf{\bibinfo{volume}{68}},
  \bibinfo{pages}{081302} (\bibinfo{year}{2003}).

\bibitem{PSSB:PSSB201350261}
\bibinfo{author}{Ganichev, S.~D.} \& \bibinfo{author}{Golub, L.~E.}
\newblock \bibinfo{title}{Interplay of {R}ashba/{D}resselhaus spin splittings
  probed by photogalvanic spectroscopy -- {A} review}.
\newblock \emph{\bibinfo{journal}{phys. status solidi (b)}}
  \textbf{\bibinfo{volume}{251}}, \bibinfo{pages}{1801--1823}
  (\bibinfo{year}{2014}).

\bibitem{Ivchenko_JETP_Letters_1989}
\bibinfo{author}{Ivchenko, E.~L.}, \bibinfo{author}{Lyanda-Geller, Y.~B.} \&
  \bibinfo{author}{Pikus, G.~E.}
\newblock \bibinfo{title}{Photocurrent in structures with quantum wells with an
  optical orientation of free carriers}.
\newblock \emph{\bibinfo{journal}{JETP Lett.}} \textbf{\bibinfo{volume}{50}},
  \bibinfo{pages}{175} (\bibinfo{year}{1989}).

\bibitem{Ivchenko_JETP_1990}
\bibinfo{author}{Ivchenko, E.~L.}, \bibinfo{author}{Lyanda-Geller, Y.~B.} \&
  \bibinfo{author}{Pikus, G.~E.}
\newblock \bibinfo{title}{Current of thermalized spin-oriented photocarriers}.
\newblock \emph{\bibinfo{journal}{Sov. Phys. JETP}}
  \textbf{\bibinfo{volume}{71}}, \bibinfo{pages}{550} (\bibinfo{year}{1990}).

\bibitem{Meier_Rashba_2007}
\bibinfo{author}{Meier, L.} \emph{et~al.}
\newblock \bibinfo{title}{Measurement of {R}ashba and {D}resselhaus spin-orbit
  magnetic fields}.
\newblock \emph{\bibinfo{journal}{Nature Phys.}} \textbf{\bibinfo{volume}{3}},
  \bibinfo{pages}{650--654} (\bibinfo{year}{2007}).

\bibitem{Kato2004_manipulation}
\bibinfo{author}{Kato, Y.}, \bibinfo{author}{Myers, R.~C.},
  \bibinfo{author}{Gossard, A.~C.} \& \bibinfo{author}{Awschalom, D.~D.}
\newblock \bibinfo{title}{Coherent spin manipulation without magnetic fields in
  strained semiconductors}.
\newblock \emph{\bibinfo{journal}{Nature}} \textbf{\bibinfo{volume}{427}},
  \bibinfo{pages}{50--53} (\bibinfo{year}{2004}).

\bibitem{PRL80_Kikkawa1998_ResonantSpinAmplificationinN-TypeGaAs}
\bibinfo{author}{Kikkawa, J.~M.} \& \bibinfo{author}{Awschalom, D.~D.}
\newblock \bibinfo{title}{{R}esonant {S}pin {A}mplification in $n$-{T}ype
  {G}a{A}s}.
\newblock \emph{\bibinfo{journal}{Phys. Rev. Lett.}}
  \textbf{\bibinfo{volume}{80}}, \bibinfo{pages}{4313--4316}
  (\bibinfo{year}{1998}).

\bibitem{2010_Schmalbuch}
\bibinfo{author}{Schmalbuch, K.} \emph{et~al.}
\newblock \bibinfo{title}{Two-dimensional optical control of electron spin
  orientation by linearly polarized light in {InGaAs}}.
\newblock \emph{\bibinfo{journal}{Phys. Rev. Lett.}}
  \textbf{\bibinfo{volume}{105}}, \bibinfo{pages}{246603}
  (\bibinfo{year}{2010}).

\end{thebibliography}


\

\textbf{Acknowledgments}\\
We thank G. G\"untherodt for helpful discussions, S. Pissinger for initial work on time-resolved photo-current measurements, and S. Kuhlen for help on the figures. This work was supported by DFG through FOR 912 and by RFBR.\\
\



\textbf{Corresponding author}\\
Correspondence to: Bernd Beschoten (bernd.beschoten@physik.rwth-aachen.de)
\end{document}